# Cold plasma treatment for biomedical applications: using aluminum foam to reduce risk while increasing efficacy

Zhitong Chen[1,2,†,*], Richard Obenchain[1,†], Richard E. Wirz[1,*]

[1] Department of Mechanical and Aerospace Engineering, University of California, Los Angeles, CA 90095, USA

[2] Institute of Biomedical and Health Engineering, Shenzhen Institute of Advanced Technology, Chinese Academy of Sciences, Shenzhen 518055, China

**ABSTRACT:** Plasma medicine is an emerging and innovative interdisciplinary research field combining biology, chemistry, physics, engineering, and medicine. However, safe clinical application of cold atmospheric plasma (CAP) technology is still a challenge. Here, we examine the use of aluminum (Al) foam with three pores-per-inch (PPI) ratings in clinical plasma applications. Al foams can filter sparks to avoid damage from high voltage discharge during surgery and efficiently deliver reactive species generated in CAP to the target. The sparks appear and plasma intensity increases at the foam/discharge interface, which just slightly increases the interface temperature without changing the interface microstructure during a 30-minute treatment. After CAP penetrated the Al foams, $N_2$, $N_2^+$, •OH, O, and He emission peaks were characterized, and the highest values appeared using Al foams with 10 PPI. CAP with and without Al foam intermediating was used to treat deionized water, and the results indicate CAP in combination with 10 PPI Al foam led to much higher ROS concentration than CAP alone. For melanoma cell experiments, CAP with and without Al foam had similar effect on cell viability after 30-second treatment, while CAP with the 10-PPI Al foam had much higher killing efficiency than CAP alone after 60-second treatment. In summary, 10-PPI Al foam can not only prevent damage to tissues resulting from high discharge voltage during the clinical surgery but also increase the delivery efficiency of reactive species generated in plasma for biomedical applications.

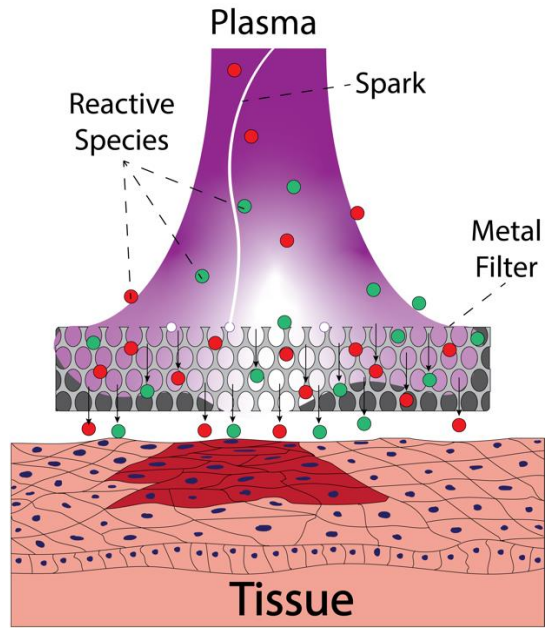

[*] E–mail address: zt.chen1@siat.ac.cn, wirz@ucla.edu

[†] These authors contributed equally: Zhitong Chen, Richard Obenchain

# ■ INTRODUCTION

Cold atmospheric plasma (CAP) has recently become a subject of great interest for a wide variety of applications including biomedicine, agriculture, catalysis, advanced materials, and so on.[1-7] All functions can be achieved due to the components of CAP: reactive oxygen species (ROS), reactive nitrogen species (RNS), free radicals, ultraviolet (UV) photons, electrons, charged particles, electromagnetic fields, and physical forces.[8-10] CAP can initiate, promote, control, and catalyze various complex behaviors and responses in biological systems. One of the main biomedical benefits of CAP is enabling selective cell death via a minimally invasive surgery.[11] CAP has opened new horizons for sterilization, coagulation, wound healing, treatment of skin infections, tissue regeneration, dental treatment, cancer therapies, and many more applications.[12-18] According to studies, the efficacy of CAP for biomedical applications relies on the synergistic action of ROS and RNS. Low doses of ROS/RNS have been reported to varyingly cause cell proliferation or cell death, while a high dose can damage lipids, proteins, and DNA and result in apoptosis.[19,20] Currently, CAP has had significant success with both in vitro and in vivo studies on cancer applications, such as lung carcinoma, breast cancer, brain cancer, melanoma, pancreatic carcinoma, and cervical carcinoma.[21-26]

One of the aims of CAP research is to utilize a plasma jet inside a living body. CAP is generally not applicable for in vivo cancer treatment in the conventional form due to high voltage and the formation of discharge in the organ.[27,28] However, some studies have employed CAP inside the body through miniaturization of the plasma to micron-sized probes.[19,29,30] Moreover, other researchers optimized the plasma devices in the micron-sized shape to deliver plasma-generated species directly into living animals.[31,32] Although micron-sized plasma devices avoid high discharge issues, they largely reduce the delivery efficiency of plasma-generated species. Thus, we propose a safe method for direct application of CAP in vivo using Al foam to filter the high voltage effects without reducing (and in some cases even increasing) plasma species delivery efficiency.

## ■ EXPERIMENTAL SECTION

**Cell lines**. The murine B16F10 melanoma cell line was purchased from UNC tissue culture facility. The B16F10-fLuc cell line was purchased from Imanis Life Sciences. Cells were cultured in DMEM (Invitrogen) with 10% FBS (Invitrogen) and 100 U/mL penicillin/streptomycin (Invitrogen) at 37 °C in 5% $CO_2$.

**Aluminum (Al) foams**. Al foams with 10 PPI (pores per inch), 20 PPI, and 40 PPI were provided by ERG Aerospace Corp. These Al foams are true metal skeletal structures, which are not a sintered, coated, or plated product. Their purity is typically that of the parent alloy metal, with no voids, inclusions, or entrapments. The matrix of cells and ligaments is completely repeatable, regular, and uniform throughout the entirety of the material.

**CAP device configuration**. The CAP device (Figure S1) was developed by employing a 3D printer (LulzBot TAZ 6) at UCLA. It consists of a two-electrode assembly with a powered needle electrode and a grounded outer ring electrode, which were connected to a high voltage transformer. Helium (He) was employed as feeding gas and its flow rate was approximately 16 L/min. The visible CAP jet had a length of approximately 5 cm and was well collimated along the entire length. The discharge voltage for the plasma jet was measured using high voltage probe and oscilloscope at approximately 8.3 kV (peak-peak) and a frequency of approximately 9.5 kHz (Figure S2). Analogous to previous studies, the plasma jet is discontinuous and represents a series of propagating plasma bundles (two bundles per HV period) with peak discharge current up to few hundred milli-amperes.

**Optical emission spectroscopy**. A fiber-coupled optical spectrometer (LR1-ASEQ Instruments), with a range of wavelength 300-1000 nm, was employed to detect CAP generated ROS and RNS (such as nitric oxide [NO], nitrogen cation [$N_2^+$], atomic oxygen [O], and hydroxyl radicals [•OH]). The optical probe was placed at a radial distance of 10 mm from the center of the target (Figure S3). Data were collected with an integration time of 10000 ms. The reactive oxygen species (ROS) and reactive nitrogen species (RNS) are shown in Figure S4.

**Thermal measurements**. The temperatures of the plasma jet and plasma jet-interacting plate and foams were made with a tripod-mounted long-wavelength infrared camera (FLIR A655sc) from a distance of approximately 15 cm diagonally above (plate) or level with the subject; the relative position of the camera to the subject remained consistent for all tests with a single subject. Frame sequences for each timepoint consisted of a multi-second exposure recorded using Research IR 4.40 with individual frames extracted as needed after recording. A linear scale manually configured from 18 °C to 30 °C was selected for all images to allow for a sufficient dynamic range.

**Diameter of foam pore**. Foam pore diameter measurements were made using a BW500 Digital Microscope to examine the surface of each sample. For each foam sample, four intact complete pores (fully bounded by ligaments in a roughly ovoid shape) were located at various positions on the surface. The longest diameter for each sample was measured in pixels and converted to millimeter length scale through calibration with a known length; the mean of the diameter measurements for each sample was then determined.

**SEM and EDX experiments**. Surface features and composition for the foam samples both prior to treatment and post-treatment were analyzed via scanning electron microscope (SEM) and energy-dispersive X-ray spectroscopy (EDS) using a ZEISS Supra 40VP SEM with a Thermo Noran System SIX EDS (courtesy of UCLA's Electron Imaging Center for NanoMachines). Prior to treatment, each sample (three Al foams of varying pore size and the metal plate) was scanned in the target area for a primary feature for inspection; the relative coordinates of the feature from the front right corner of the sample were noted, and four images at increasing resolution were taken. For each resolution, EDS results were also obtained. Post-treatment, the same features were located using the previously determined coordinates, and images and EDS analyses were performed at the same magnification levels as pre-treatment.

**ROS measurement**. A fluorimetric hydrogen peroxide assay Kit (Sigma-Aldrich) was used for measuring the amount of $H_2O_2$ according to the manufacturer's protocol. Briefly, 50 $\mu$l of standard curve, control, and experimental samples were added to 96-well flat-bottom black plates, and then 50 $\mu$l of Master Mix was added to each of well. The plates were incubated for 20 min at room temperature protected from light and fluorescence was measured by a Tecan Infinite® 200 PRO Multimode Plate Reader at Ex/Em: 540/590 nm.

**Cell viability**. The cells were plated in 6-well flat-bottomed microplates at a density of 500000 cells per well in 1.5 mL of complete culture medium. Cells were incubated for 24 hours to ensure proper cell adherence and stability. On day 2, plasma jet and plasma penetrating different PPI foams treated melanoma cells without cultured medium. Following treatment, 1.5 mL cultured medium was added to each well, and cells were further incubated at 37 °C for 24 hours. The viability of the melanoma cells was measured with a Cell Counting Kit (CCK) 8 assay. 100 $\mu$L of CCK 8 reagent was added per well. The plates were incubated for 3 hours at 37 °C. The absorbance was measured at 450 nm using a Tecan Infinite® 200 PRO Multimode Plate Reader.

# ■ RESULTS AND DISCUSSION

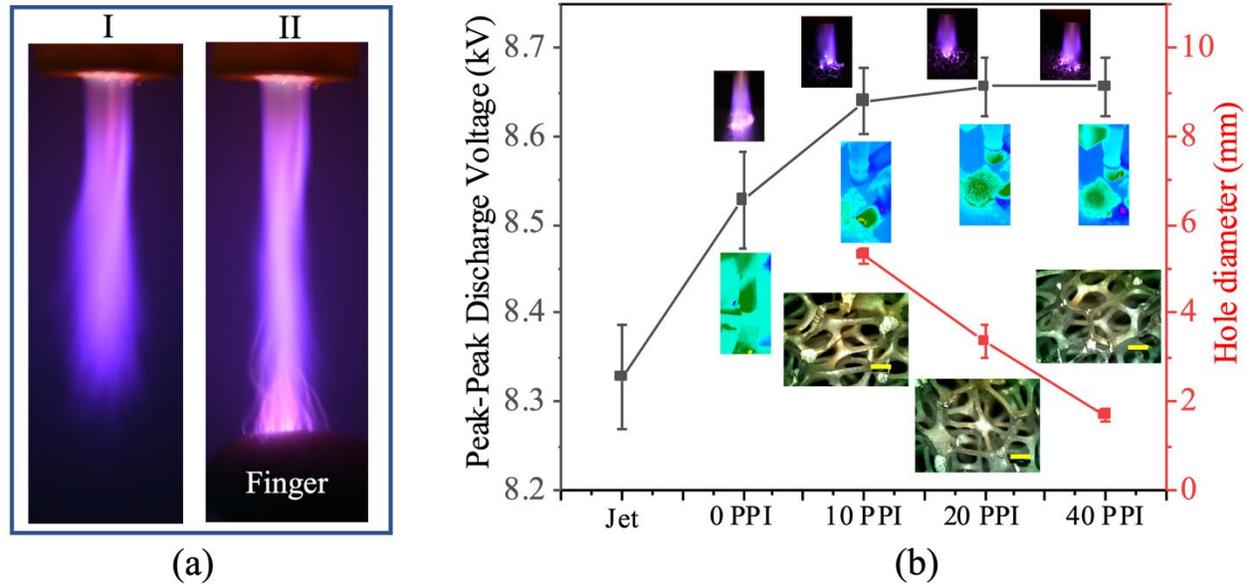

**Figure 1.** (a) (I) A typical plasma jet and (II) sparks appear when a finger is close to the jet. (b) Diameters and structures of 10 PPI, 20 PPI, and 40 PPI Al foams. Different PPI affects plasma discharge voltage and temperature distribution.

A typical image of the CAP jet is shown in Figure 1a (I) for helium feeding corresponding to the flow rate about 16 L/min. The visible CAP jet had a length of approximately 5 cm and was well collimated along the entire length. As a person's finger closes to the tip of plasma jet, the individual feels a slight stinging, and sparks appear (Figure 1a(II)). The stinging and sparks become stronger when distance is decreased. These sparks are dangerous to sensitive organs of living body during plasma use for clinical surgery. The Al foams with 10 PPI, 20 PPI, and 40 PPI are shown in Figure 10b, and the average diameter of pores for each PPI are 5.3±0.2 mm, 3.4±0.4 mm, and 1.7±0.2 mm, respectively. Here, we define Al plate as 0 PPI. When the plasma jet interacts with Al foams (0 PPI, 10 PPI, 20 PPI, and 40 PPI), the peak-peak discharge voltage increased (Figure 1b). While Al foams (10 PPI, 20 PPI, and 30 PPI) lead to higher discharge voltages than the Al plate (0 PPI), Al foams with 10 PPI, 20 PPI, and 40 PPI have similar discharge voltages. This increased discharge voltage and current when the plasma jet interacts with Aluminum foams increases the background electric field in the space of CAP jet propagation, leading to increased electron density.[33] We also investigated the effect of the plasma jet on the temperature at the interface of Al foams with different PPI, where we found a slight thermal impact on Al foams (Figure 1b): the temperature at the interface of 10 PPI, 20 PPI, and 40 PPI Al foams increased by approximately 5.4%, 4.8%, and -0.4% after 30-minute treatment, respectively (Figure S5).

The interaction of a plasma discharge with a material surface, either porous or solid, has been most comprehensively studied in relation to plasma sputtering.[34-36] For CAP, low ion and neutral energies imply that all relevant interaction between the plasma and any material is mediated by electrons, leading to a negatively-charged plasma sheath with constant interaction between the sheath and the plasma to maintain a quasineutral plasma state[37]. Sheath modeling for CAP has generally been performed in relation to the electrodes, with the sheath at the grounded electrode exhibiting a smaller thickness than that at the charged electrode. The electrical interaction between target and plasma jet when the target is conductive and placed within the plume results not only in changes to the electrical characteristics of the plasma jet in operation but also can lead to sporadic arcing between the target and the electrodes within the jet itself.[38] In the current configuration, the Al foam is placed within the plume between the plasma and the final target. This arrangement protects the final target from any stray electrical effects that could occur through interaction with the plume without increasing the physical distance between the discharge and the target. Inclusion of the foam also led to an increased discharge current from the power source, consistent with modeling the ungrounded conductive foam as a parallel resistance within the circuit. For a constant potential difference at the electrodes, the corresponding power input thus increased in comparison to operation in the absence of the foam. Even without specific electron temperature or density measurements, the Debye length $\lambda_d$ for a He plasma jet has been shown to be on the order of 1-10 microns.[39] The resulting Child-Langmuir ion sheath $L_S$ can then be approximated as up to $20\lambda_d$, which even at upper ranges remains much smaller than the pore diameter for the densest foam sample ($L_S$ <0.2mm, as compared to 1.5mm for the 40 PPI sample). Thus, the increased discharge current is likely explained as inclusion of the gaps within the foam in plasma generation, leading to increased reactive species formation.

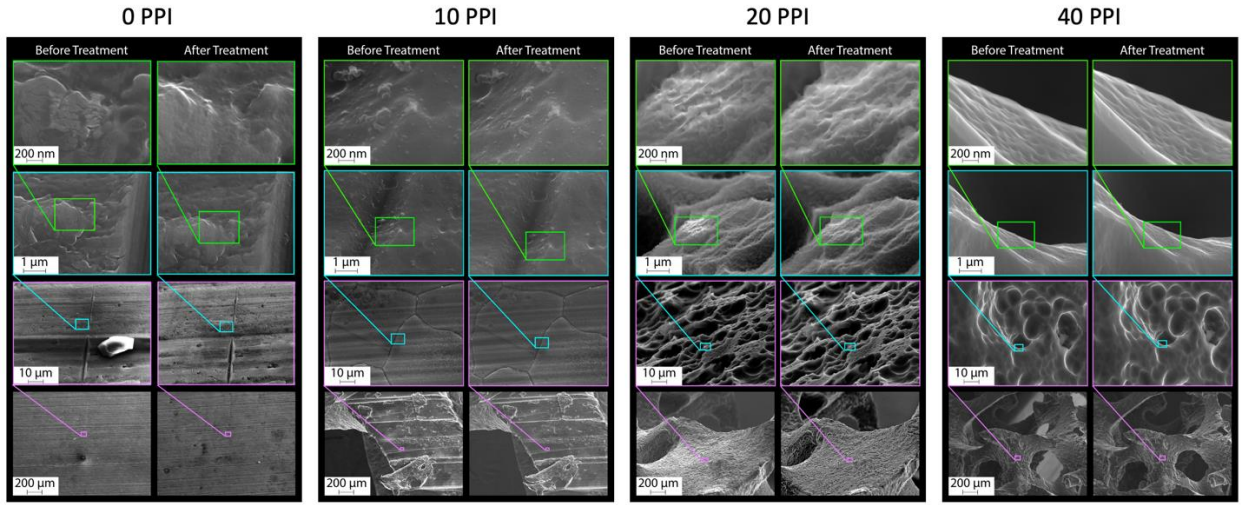

**Figure 2**. Microstructure surface morphology of Al foams with 10 PPI, 20 PPI, and 40 PPI before and after 30-minute plasma jet interactions.

The possibility of surface deformation due to plasma jet exposure was examined using visual inspection of SEM imaging at multiple magnifications. Images were taken of each foam subject at the 200 μm, 10 μm, 1 μm, and 200 nm scales both prior to exposure and post-exposure (Figure 2). At the 200 and 10 μm scales, no major changes are noticed in any sample beyond minor debris addition or removal. At 1 μm, the 20 PPI and 40 PPI samples showed no changes at all, while the 10 PPI sample revealed minor changes of surface artifacts in the 200 nm image that appear to be the result of changes in target angle within the SEM rather than deformation. EDX data shows that elements of Al foams do not change after plasma treatment, for example, the element area profile of 10 PPI Al foams before and after plasma treatment is shown in Figure S6. At 1 μm, the 0 PPI (solid plate) sample reveals contour changes, which become more obvious at the 200 nm scale.

For example, the sharply-defined ridge starting in the upper-right of the before-treatment image (Figure 2) becomes amorphous, with multiple layers of curves in place of the original structures. Additionally, on the upper right, a similar bi-lobed curve has been deformed drastically into a rough corner in the post-treatment image. These contour changes cannot be accounted for by shadowing or other imaging effects and are likely the result of surface melting under exposure to the plasma. While the bulk temperature of the plate never exceeded 30°C, localized melting as a result of arcing could explain such deformation.

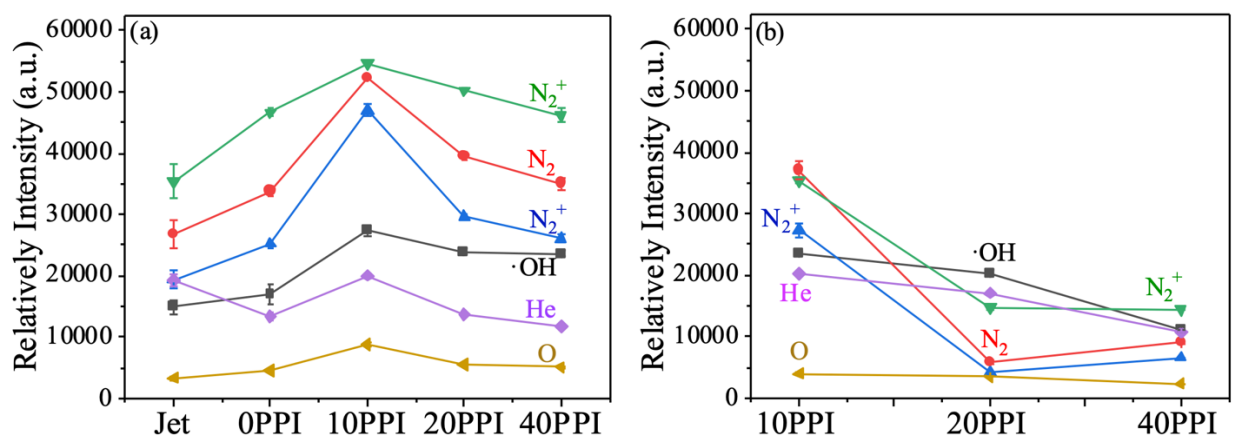

Figure 3. (a) The effect of Al foams with 0 PPI, 10 PPI, 20 PPI, and 40 PPI on reactive species of plasma jet. (b) Reactive species of plasma jet penetrating Al foams with 10 PPI, 20 PPI, and 40 PPI.

The optical emission distributions of reactive species generated by plasma before and after penetrating Al foams with different PPI are shown in Figure 3, Figure S7, and Figure S8. The identification of emission lines and bands were performed according to reference[40]. The Helium bands were assigned at 588, 668, and 705 nm. We anticipated that •OH was present at 309 nm. The wavelength of 337, 376, and 381 nm could be indicative of the low-intensity $N_2$ second-positive system ($C^3\Pi_u$-$B^3\Pi_g$). In addition, their magnitudes are at most a few thousandths of the highest peak $N_2^+$ (391 nm), also 358 and 428 nm are $N_2^+$. It is shown that O (777 nm), •OH (309 nm), $N_2^+$ (391 nm), and $N_2$ (337nm) are the domain species of the spectra. For $N_2$, $N_2^+$, •OH, O, and He emission peaks between electrodes and foams, their relatively intensities appear highest at 10 PPI with the exception of He during plasma-foams interactions (Figure 3a). After penetrating Al foams, the highest intensity of these six emission peaks still appears with the 10 PPI Al foams (Figure 3b). The reason for this should be the larger diameter of pores for 10 PPI than 20 PPI and 40 PPI Al foams. As the EM field generated by the electrodes fluctuates, it is likely that the sheaths surrounding the ligaments in the foam fluctuate within a range of sizes. The remaining space within the pores of the foam allows for reactive species and at least partial ion passage through the foam to the target. The intensity of •OH, O, and He emission peaks reduce with the diameter of Al foam pore decreasing, while the intensity of $N_2$ and $N_2^+$ emission peaks appear lowest at 20 PPI Al foam, which are slightly lower than 10 PPI. The floating charge on the foam itself due to electrical interaction with the plume as well as potential infusion of the foam with the plasma discharge likely modifies the flow of charged species through the foam, but further analysis is necessary to determine the exact nature of the interaction.

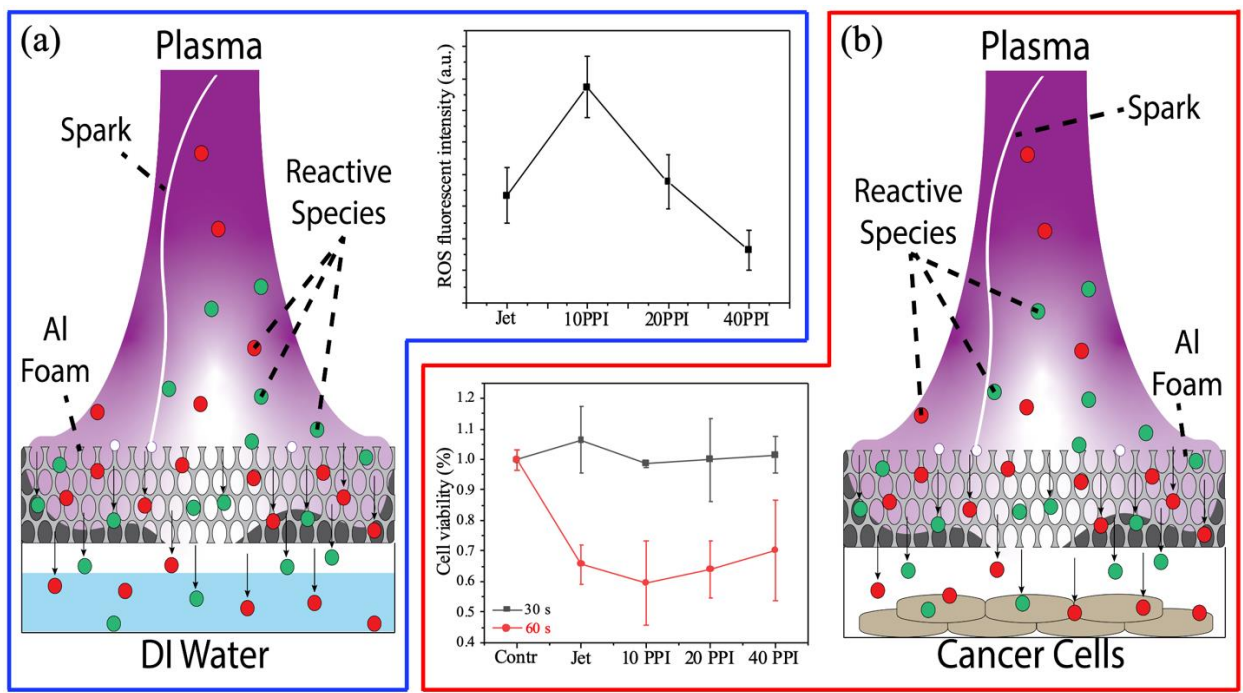

**Figure 4**. Reactive species produced in plasma penetrated Al foams with 10 PPI, 20 PPI, and 40 PPI to react with DI water (2 mL) and generate ROS in the DI water (treatment time: 60 seconds). (b) Plasma-generated ROS/RNS penetrated Al foams to interact with melanoma cells (treatment time: 30 and 60 seconds) and kill them (cell viability measured after 24 hours).

Figure 4 shows Al foams filtering sparks and reactive species penetrate Al foams to reach deionized (DI) water and cancer cells. Compared to the plasma jet without Al foams, the resulting ROS concentration in DI water after filtering using 10 PPI Al foam is much higher (Figure 4a). The ROS concentration with 20 PPI Al foam is slightly higher than plasma jet, while 40 PPI is much lower than plasma jet. The reactive species from the water interface transfer into the water to generate $H_2O_2$ and $NO_2^-$, and the reaction mechanisms at the water interface and in the water can found in our previous publications.[41-43] The efficacy of plasma for cancer therapy relies on the synergistic action of reactive species. For example, active oxygen might promote a "plasma killing effect", while nitric oxide could produce a "plasma healing" effect.[44] Figure 4b shows reactive species from the plasma jet alone and plasma jet penetrating Al foams with 10 PPI, 20 PPI, and 40 PPI interact with melanoma cells and induce melanoma cells proliferation/death after 30-second and 60-second duration. 30-second treatment seems not to kill cells while slightly promoting cell proliferation due to the low dose of reactive species. When the treatment time increases to 60 seconds, some melanoma cells are killed, and 10 PPI Al foams exhibit the best killing efficiency. The results from Figure 3 and Figure 4a also support this finding. High doses of reactive species can induce cell death through damage to lipids, DNA, and proteins.[45,46] For example, •OH itself and •OH-driven amino peroxides can react with DNA, causing various forms of damage.[47] O and other ROS not only generate oxidative damage in many biological targets but also change mitochondrial membrane potentials, resulting in cell death.[48-50] Moreover, $H_2O_2$ and $NO_2^-$ easily form ONOO$^-$ once they collide, and ONOO$^-$ is a powerful oxidant and nitrating agent highly damaging to tumor cells.[51] Al foam with 10 PPI not only filters the sparks but increases the delivery of reactive species for biomedical applications.

■ **CONCLUSIONS**

In summary, this work demonstrates a novel and safe method for deploying CAP therapeutically in vivo while avoiding potential high voltage-related tissue damage through the use of Al foams as filter. Al foams with 0 PPI, 10 PPI, 20 PPI, and 40 PPI increase the discharge voltage of CAP jet. Discharge-related sparks appear, and plasma intensity increases at the Al foam interface, which slightly increases the interface temperature without changing the interface microstructure during a 30-minute treatment. Once CAP penetrates the Al foams, $N_2$, $N_2^+$, •OH, O, and He emission peaks exhibit the highest values for Al foams with 10 PPI. Compared to the unfiltered CAP jet, plasma penetrating 10 PPI Al foam to react with DI water generates much higher ROS concentration. Filtering with 10 PPI foam also reveals higher killing efficiency for melanoma cells than the unfiltered CAP jet after 60-second treatment. Al foam with 10 PPI not only prevents potential damage to sensitive tissues from the high discharge voltage during clinical surgery, but also increases the delivery efficiency of reactive species generated in the plasma for biomedical applications. This method expands the potential uses for CAP during clinical surgery and is also a significant milestone for employing CAP towards direct tumor treatment within a living patient.

**Acknowledgement**

This work was supported by a grant from the Air Force Office of Scientific Research (FA9550-14-10317, UCLA Subaward No. 60796566-114411, to R.W.), the National Key R&D Program of China (2022YFE0126000, to Z.C.), and the Guangdong Basic and Applied Basic Research Foundation (2022A1515011129, to Z.C.).

# References

1. Chen, Z. & Wirz, R. E. Cold Atmospheric Plasma (CAP) Technology and Applications. *Synthesis Lectures on Mechanical Engineering* **6**, i-191 (2021).
2. Chen, G. *et al.* Transdermal cold atmospheric plasma-mediated immune checkpoint blockade therapy. *Proceedings of the National Academy of Sciences* **117**, 3687-3692 (2020).
3. Lu, X. *et al.* Reactive species in non-equilibrium atmospheric-pressure plasmas: Generation, transport, and biological effects. *Physics Reports* **630**, 1-84 (2016).
4. Bourke, P., Ziuzina, D., Boehm, D., Cullen, P. J. & Keener, K. The potential of cold plasma for safe and sustainable food production. *Trends in biotechnology* **36**, 615-626 (2018).
5. Mehta, P. *et al.* Overcoming ammonia synthesis scaling relations with plasma-enabled catalysis. *Nature Catalysis* **1**, 269-275 (2018).
6. Neyts, E. C., Ostrikov, K., Sunkara, M. K. & Bogaerts, A. Plasma catalysis: synergistic effects at the nanoscale. *Chemical reviews* **115**, 13408-13446 (2015).
7. Dou, S. *et al.* Plasma‐Assisted Synthesis and Surface Modification of Electrode Materials for Renewable Energy. *Advanced materials* **30**, 1705850 (2018).
8. Gjika, E. *et al.* Adaptation of operational parameters of cold atmospheric plasma for in vitro treatment of cancer cells. *ACS applied materials & interfaces* **10**, 9269-9279 (2018).
9. Li, M. *et al.* Advances in plasma-assisted ignition and combustion for combustors of aerospace engines. *Aerospace Science and Technology* **117**, 106952 (2021).
10. Zheng, H., Chen, P. & Chen, Z. Cold-atmospheric-plasma–induced skin wrinkle. *EPL (Europhysics Letters)* **133**, 15001 (2021).
11. Lu, X. *et al.* Transcutaneous plasma stress: From soft-matter models to living tissues. *Materials Science and Engineering: R: Reports* **138**, 36-59 (2019).
12. Weltmann, K. D. *et al.* Atmospheric-pressure plasma sources: Prospective tools for plasma medicine. *Pure and Applied Chemistry* **82**, 1223-1237 (2010).
13. Chen, Z., Lin, L., Cheng, X., Gjika, E. & Keidar, M. Treatment of gastric cancer cells with nonthermal atmospheric plasma generated in water. *Biointerphases* **11**, 031010 (2016).
14. Isbary, G. *et al.* Successful and safe use of 2 min cold atmospheric argon plasma in chronic wounds: results of a randomized controlled trial. *British Journal of Dermatology* **167**, 404-410 (2012).
15. Chen, Z., Zhang, S., Levchenko, I., Beilis, I. I. & Keidar, M. In vitro demonstration of cancer inhibiting properties from stratified self-organized plasma-liquid interface. *Scientific reports* **7**, 1-11 (2017).
16. Chen, G. *et al.* Portable air-fed cold atmospheric plasma device for postsurgical cancer treatment. *Science advances* **7**, eabg5686 (2021).
17. Chen, Z., Garcia Jr, G., Arumugaswami, V. & Wirz, R. E. Cold atmospheric plasma for SARS-CoV-2 inactivation. *Physics of Fluids* **32**, 111702 (2020).
18. Chen, Z., Xu, R.-G., Chen, P. & Wang, Q. Potential agricultural and biomedical applications of cold atmospheric plasma-activated liquids with self-organized patterns formed at the interface. *IEEE Transactions on Plasma Science* **48**, 3455-3471 (2020).
19. Chen, Z. *et al.* A novel micro cold atmospheric plasma device for glioblastoma both in vitro and in vivo. *Cancers* **9**, 61 (2017).

20  Keidar, M. Plasma for cancer treatment. *Plasma Sources Science and Technology* **24**, 033001 (2015).
21  Li, W. *et al.* Cold atmospheric plasma and iron oxide-based magnetic nanoparticles for synergetic lung cancer therapy. *Free Radical Biology and Medicine* **130**, 71-81 (2019).
22  Xiang, L., Xu, X., Zhang, S., Cai, D. & Dai, X. Cold atmospheric plasma conveys selectivity on triple negative breast cancer cells both in vitro and in vivo. *Free Radical Biology and Medicine* **124**, 205-213 (2018).
23  Akter, M., Jangra, A., Choi, S. A., Choi, E. H. & Han, I. Non-Thermal Atmospheric Pressure Bio-Compatible Plasma Stimulates Apoptosis via p38/MAPK Mechanism in U87 Malignant Glioblastoma. *Cancers* **12**, 245 (2020).
24  Adhikari, M. *et al.* Cold atmospheric plasma and silymarin nanoemulsion synergistically inhibits human melanoma tumorigenesis via targeting HGF/c-MET downstream pathway. *Cell Communication and Signaling* **17**, 52 (2019).
25  Verloy, R., Privat-Maldonado, A., Smits, E. & Bogaerts, A. Cold Atmospheric Plasma Treatment for Pancreatic Cancer—The Importance of Pancreatic Stellate Cells. *Cancers* **12**, 2782 (2020).
26  Ahn, H. J. *et al.* Atmospheric-pressure plasma jet induces apoptosis involving mitochondria via generation of free radicals. *PloS one* **6**, e28154 (2011).
27  Mirpour, S. *et al.* Utilizing the micron sized non-thermal atmospheric pressure plasma inside the animal body for the tumor treatment application. *Scientific reports* **6**, 29048 (2016).
28  Chen, Z., Obenchain, R. & Wirz, R. E. Tiny cold atmospheric plasma jet for biomedical applications. *Processes* **9**, 249 (2021).
29  Kim, J. Y. *et al.* Single‐cell‐level microplasma cancer therapy. *Small* **7**, 2291-2295 (2011).
30  Chen, Z. *et al.* Micro-sized cold atmospheric plasma source for brain and breast cancer treatment. *Plasma Medicine* **8** (2018).
31  Zuo, X., Wei, Y., Wei Chen, L., Dong Meng, Y. & Team, P. M. Non-equilibrium atmospheric pressure microplasma jet: An approach to endoscopic therapies. *Physics of Plasmas* **20**, 083507 (2013).
32  Sohbatzadeh, F. & Omran, A. V. The effect of voltage waveform and tube diameter on transporting cold plasma strings through a flexible dielectric tube. *Physics of Plasmas* **21**, 113510 (2014).
33  Lin, L. & Keidar, M. Cold atmospheric plasma jet in an axial DC electric field. *Physics of Plasmas* **23**, 083529 (2016).
34  Ottaviano, A. *et al.* Plasma-Material Interactions for Electric Propulsion: Challenges, Approaches and Future.  (2019).
35  Li, G. Z. *et al.* In situ plasma sputtering and angular distribution measurements for structured molybdenum surfaces. *Plasma Sources Science and Technology* **26**, 065002 (2017).
36  Ottaviano, A. *et al.* In situ microscopy for plasma erosion of complex surfaces. *Review of Scientific Instruments* **92**, 073701 (2021).
37  Cooper, M. A., Holle, R. & Lopez, R. Recommendations for lightning safety. *JAMA* **282**, 1132-1133 (1999).

38  Judée, F., Vaquero, J., Guégan, S., Fouassier, L. & Dufour, T. Atmospheric pressure plasma jets applied to cancerology: correlating electrical configuration with in vivo toxicity and therapeutic efficiency. *Journal of physics D: Applied physics* **52**, 245201 (2019).

39  Jiang, C., Miles, J., Hornef, J., Carter, C. & Adams, S. Electron densities and temperatures of an atmospheric-pressure nanosecond pulsed helium plasma jet in air. *Plasma Sources Science and Technology* **28**, 085009 (2019).

40  Pearse, R. W. B. & Gaydon, A. G. *Identification of molecular spectra*.  (Chapman and Hall, 1976).

41  Chen, Z., Cheng, X., Lin, L. & Keidar, M. Cold atmospheric plasma discharged in water and its potential use in cancer therapy. *Journal of Physics D: Applied Physics* **50**, 015208 (2016).

42  Chen, Z., Lin, L., Cheng, X., Gjika, E. & Keidar, M. Effects of cold atmospheric plasma generated in deionized water in cell cancer therapy. *Plasma Processes and Polymers* **13**, 1151-1156 (2016).

43  Chen, Z. *et al.* Cold atmospheric plasma delivery for biomedical applications. *Materials Today* **54**, 153-188 (2022).

44  Keidar, M. A prospectus on innovations in the plasma treatment of cancer. *Physics of Plasmas* **25**, 083504 (2018).

45  Hong, Y., Zeng, J., Wang, X., Drlica, K. & Zhao, X. Post-stress bacterial cell death mediated by reactive oxygen species. *Proceedings of the National Academy of Sciences* **116**, 10064-10071 (2019).

46  Redza-Dutordoir, M. & Averill-Bates, D. A. Activation of apoptosis signalling pathways by reactive oxygen species. *Biochimica et Biophysica Acta (BBA)-Molecular Cell Research* **1863**, 2977-2992 (2016).

47  Gebicki, S. & Gebicki, J. M. Crosslinking of DNA and proteins induced by protein hydroperoxides. *Biochemical Journal* **338**, 629-636 (1999).

48  Kos, S. *et al.* Safety aspects of atmospheric pressure helium plasma jet operation on skin: In vivo study on mouse skin. *PloS one* **12**, e0174966 (2017).

49  Simon, H.-U., Haj-Yehia, A. & Levi-Schaffer, F. Role of reactive oxygen species (ROS) in apoptosis induction. *Apoptosis* **5**, 415-418 (2000).

50  Johnson, T. M., Yu, Z.-X., Ferrans, V. J., Lowenstein, R. A. & Finkel, T. Reactive oxygen species are downstream mediators of p53-dependent apoptosis. *Proceedings of the National Academy of Sciences* **93**, 11848-11852 (1996).

51  Pacher, P., Beckman, J. S. & Liaudet, L. Nitric oxide and peroxynitrite in health and disease. *Physiological reviews* **87**, 315-424 (2007).

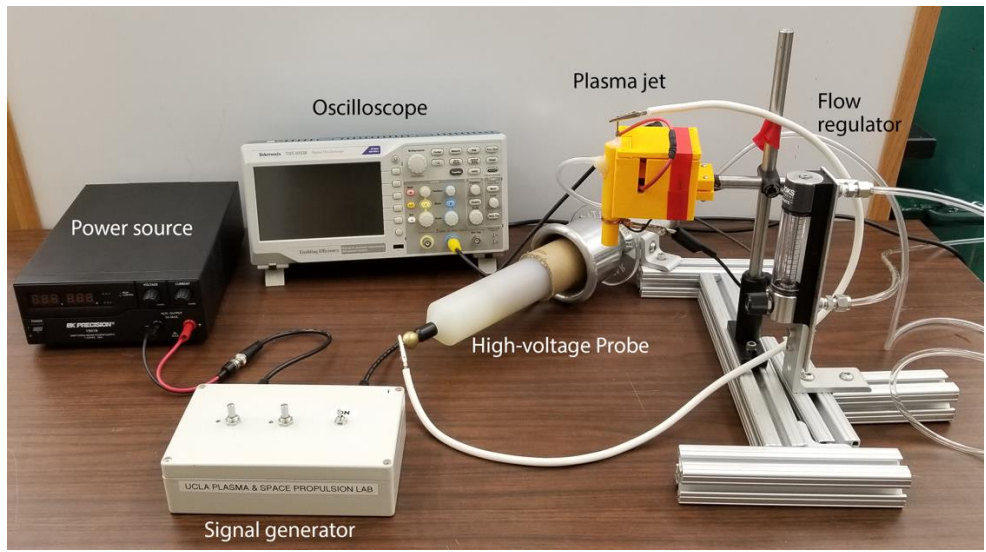

Figure S1. A schematic diagram of the experimental setup.

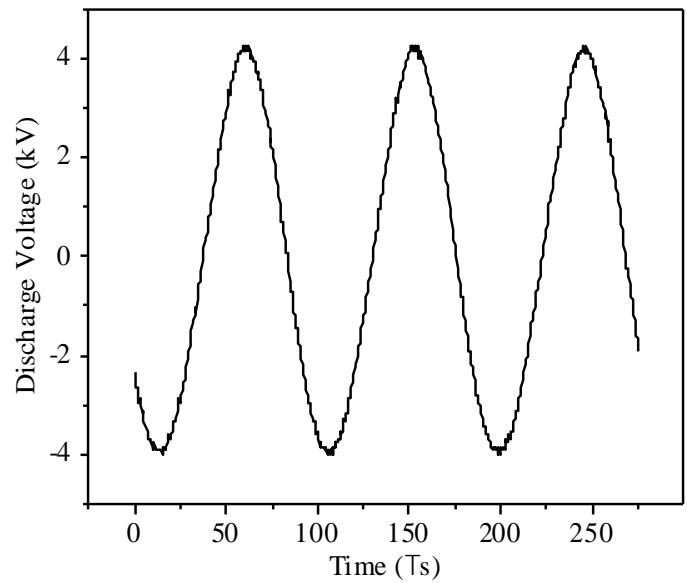

Figure S2. The discharge voltage of He plasma jet.

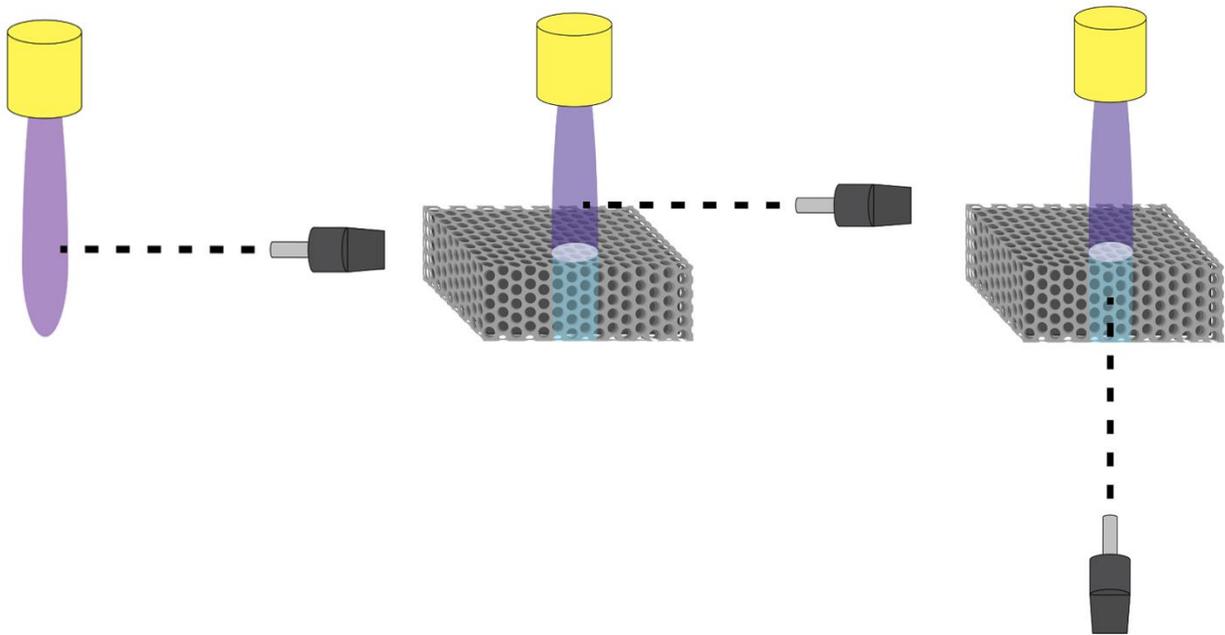

Figure S3. OES detector position for measuring plasma jet, plasma jet-foams interactions, and plasma jet penetrating foams.

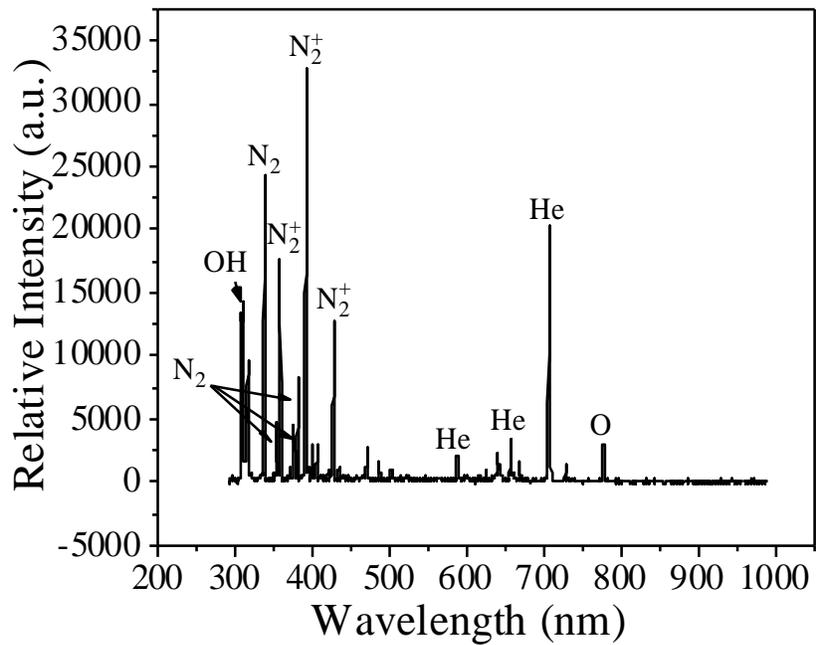

Figure S4. Optical emission spectrum detected from the He plasma jet using A fiber-coupled optical spectrometer (LR1-ASEQ Instruments), with a range of wavelength 300-1000 nm.

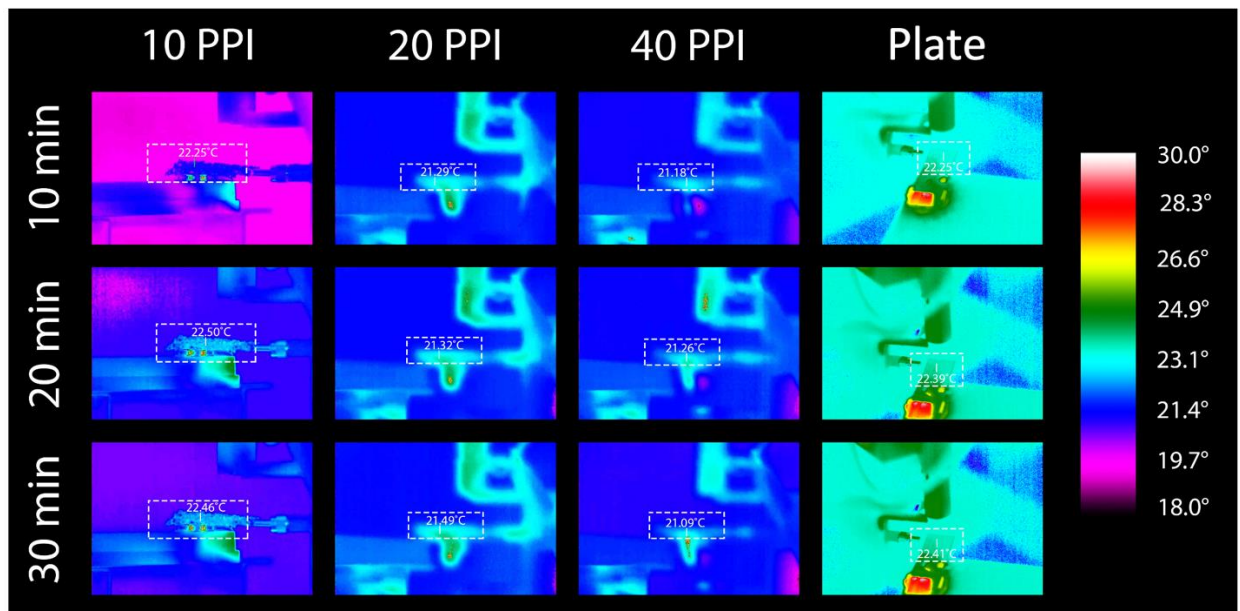

Figure S5. Temperature change of aluminum foams (10 PPI, 20 PPI, and 40 PPI) and plate with different treatment time.

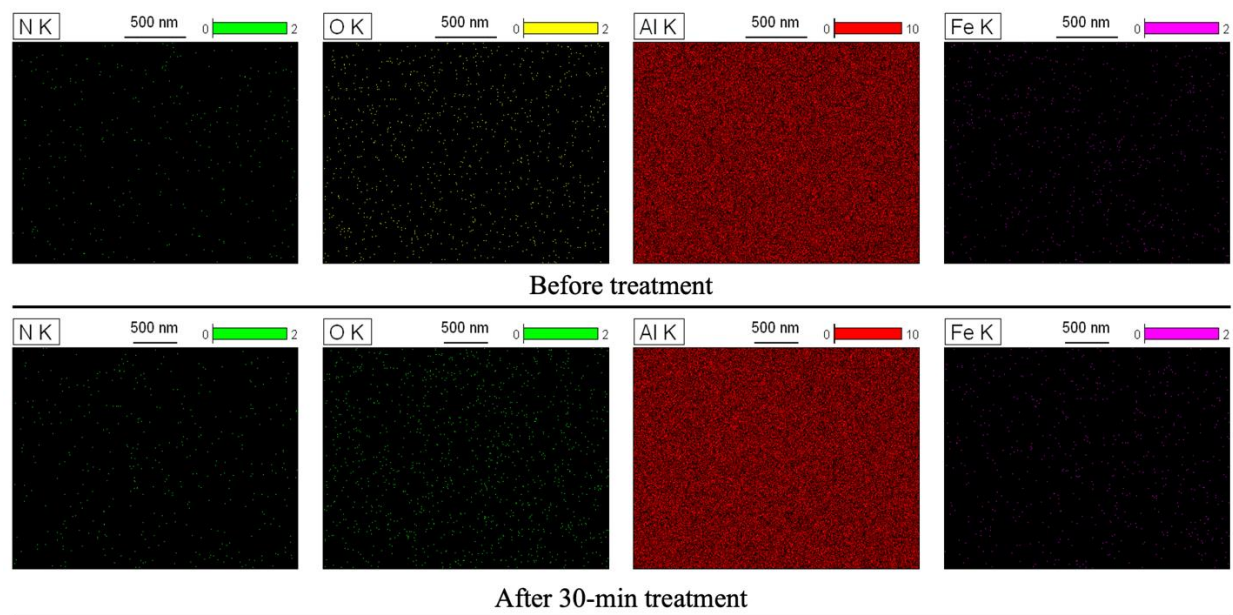

Figure S6. The element area profile of plasma treating aluminum foams with 10 PPI before and after 30 minutes.

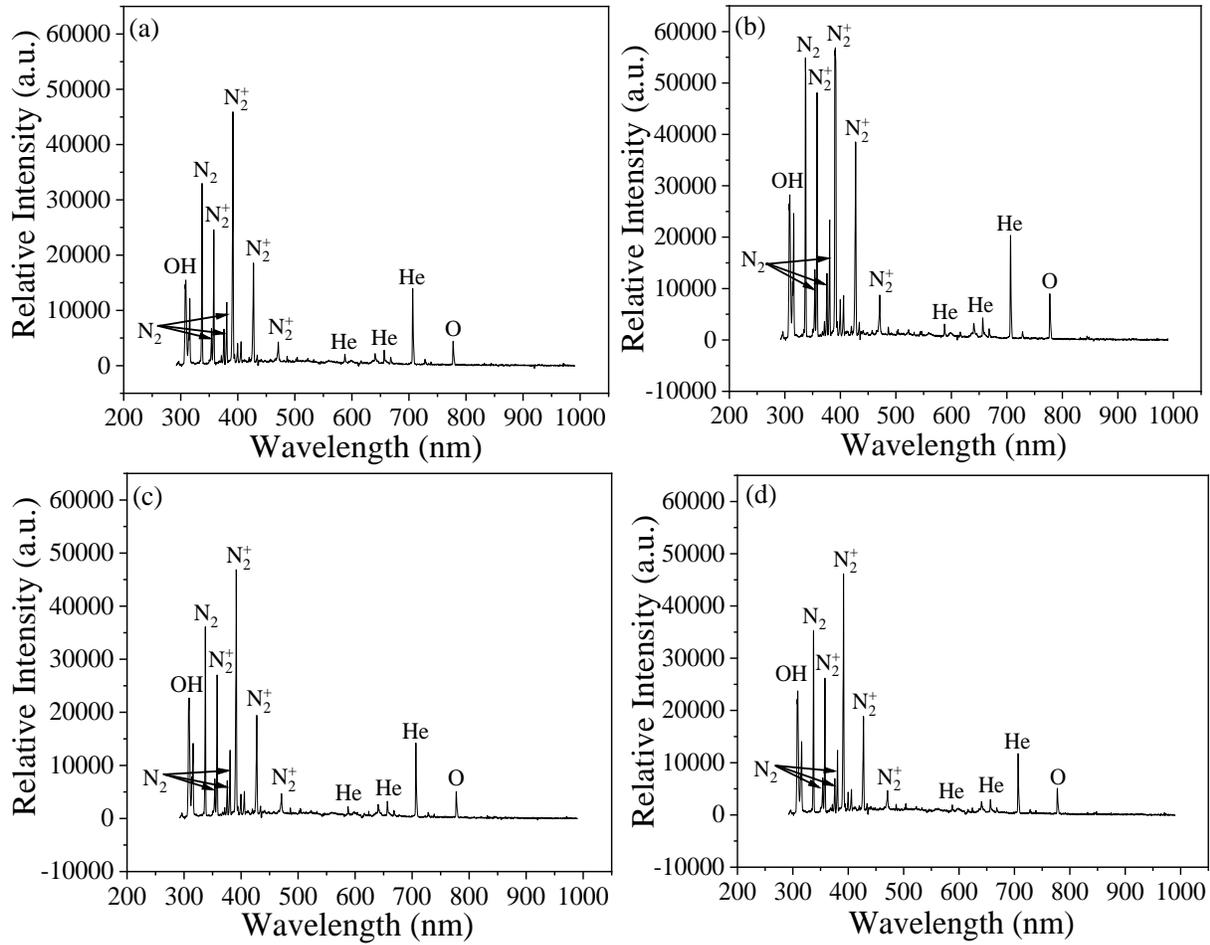

Figure S7. OES of plasma jet interacting with aluminum plate and foams: (a) plate, (b) 10 PPI, (c) 20 PPI, and (d) 40 PPI.

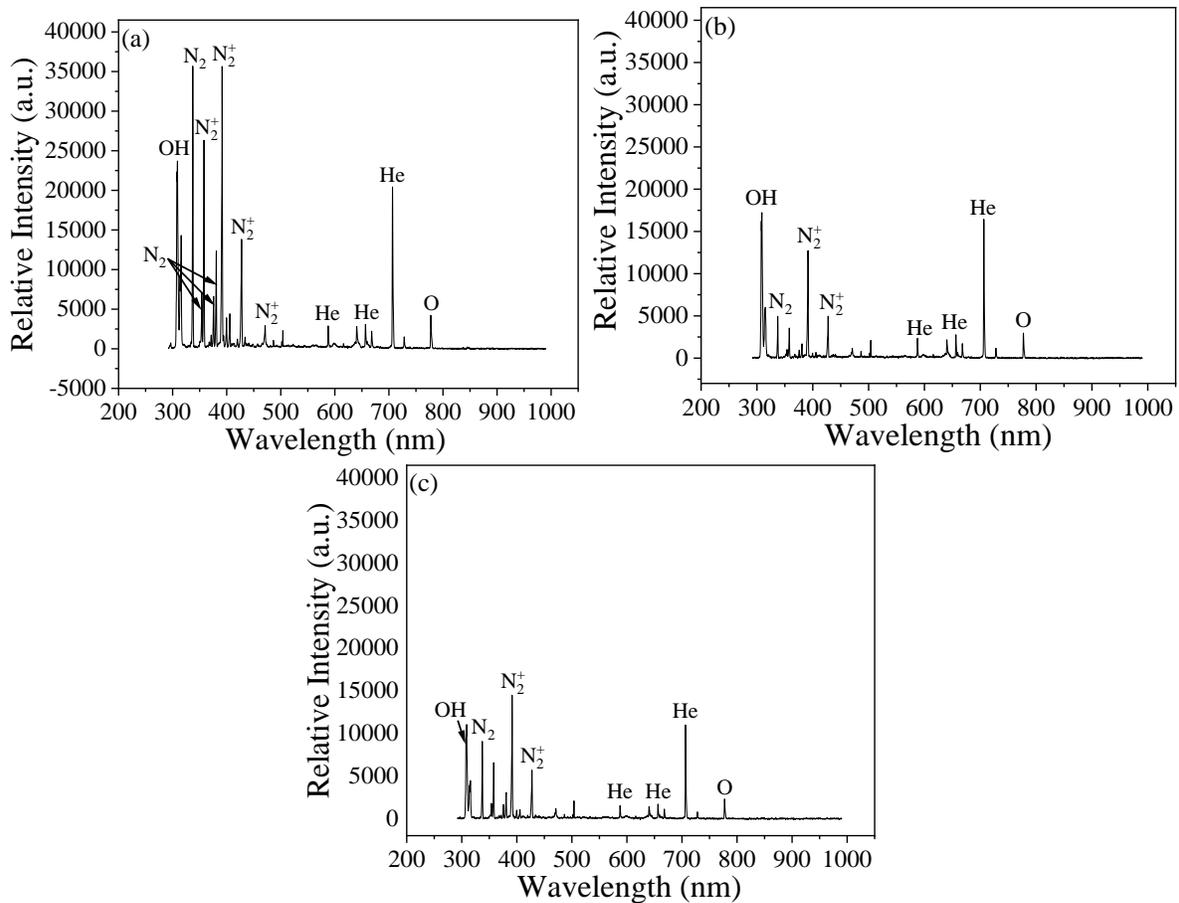
Figure S8. OES of plasma penetrating foams with different PPI: (a) 10 PPI, (b) 20 PPI, and (3) 40 PPI.